%% file: ICT_HFT_M4_CommunicationsMagazine_Final_OneFile.tex
\documentclass[12pt,comsoc]{IEEEtran}

\usepackage[T1]{fontenc} 
\usepackage[utf8]{inputenc}
\IEEEoverridecommandlockouts
\usepackage{amsmath}
\usepackage[cmintegrals]{newtxmath}
\usepackage{bm} 
\usepackage{siunitx}
\usepackage{tikz}
\usepackage{pgfplots}
\usepackage[final]{changes}
\definechangesauthor[name={Peitzmeier}, color=red]{NP}

\pgfplotsset{compat=1.6}
\usetikzlibrary{external}
\usetikzlibrary{shapes, arrows,graphs}
\tikzset{external/force remake}
\usepackage{mathrsfs}
\graphicspath{{./pic/}}
\usepackage{soul} 

	\title{On the Feasibility of Multi-Mode Antennas in UWB and IoT Applications below 10~GHz
		}

	\author{\IEEEauthorblockN{Nils L. Johannsen,\thanks{Nils L. Johannsen and Peter A. Hoeher are with the \textit{University of Kiel};} \thanks{Nikolai Peitzmeier and Dirk Manteuffel are with the \textit{Leibniz-University Hannover}.}Nikolai Peitzmeier, Peter A. Hoeher, and Dirk Manteuffel\\}	
}
\makeatletter
\newcommand{\currentfsize}{\f@size pt}
\newcommand{\showfont}{encoding: \f@encoding{},
	family: \f@family{},
	series: \f@series{},
	shape: \f@shape{},
	size: \f@size{}
}
\makeatother
\begin{document}

	\pgfplotsset{compat=1.3,
		every axis legend/.style={
			y tick label style={/pgf/number format/1000 sep=},					
			x tick label style={/pgf/number format/1000 sep=},
			z tick label style={/pgf/number format/1000 sep=}
	}}
	
	\maketitle


	\begin{abstract}
		While on the one hand 5G and beyond 5G networks are challenged by ultra-high data rates in wideband applications like 100+ Gb/s wireless Internet access, on the other hand they are expected to support reliable low-latency Internet of Things applications with ultra-high connectivity. 
		These conflicting challenges are addressed in a system proposal dealing with both extremes. 
		In contrast to most recent publications, focus is on the frequency domain below 10~GHz. 
		Towards this goal, multi-mode antenna technology is used and different realizations, offering up to eight uncorrelated ports per radiator element, are studied.
		Possible baseband architectures tailored to multi-mode antennas are discussed, enabling different options regarding precoding and beamforming.
	\end{abstract}

	
	\section{Introduction}
	
	The upcoming Internet of Things (IoT) is expected to deliver a tremendous growth in the number of mobile devices and mobile terminals, each requiring Internet access. 
	As an example, in \cite{Militano2015} and the sources therein, about 10 billion devices till 2020 are expected, while in \cite{Ejaz2016} already a number of 20 to 40 billion IoT devices is mentioned. 
	It is commonly assumed that next generation mobile networks will play a crucial role in handling both high data rates to users streaming videos on the one hand and connecting a large number of low data rate devices, e.g. connected sensors, on the other hand. 
	As suggested in \cite{BLee2018}, rather small cells employing massive multiple-input multiple-output (MIMO) arrays are promising to handle the requirements due to the large gain and channel hardening aspects which help to reduce transmission power. 
	
	However, while connecting a large number of low data rate devices, the base station (BS) still needs to provide high data rates to individual customers. 
	In recent publications towards IoT, mmWaves (30-300~GHz), submmWaves (300-1000~GHz) or even THz signals are targeted. Due to the strong free-space attenuation, sparse scattering occurs and line-of-sight (LOS) beamforming is mandatory.
	Opposed to the mmWave approach, in this paper focus is on data rates above 100~Gb/s as well as robust IoT applications in the frequency regime below 10~GHz. 
	
	Towards these goals, multi-mode antenna technology is studied in this contribution. Multi-mode antennas behave like antenna arrays, both at BS and mobile terminal. 
	Since multiple ports per element are available, which provide orthogonal characteristic far fields, the array can be modeled by overlayed uniform planar arrays, each employing one specific antenna type at the same position. 
	Because of their orthogonal radiation patterns, a diversity gain and/or multiplexing gain is resolvable. 
	For use cases such as ultra-fast wireless Internet access, the different radiation patterns enable multi-stream processing. 
	Whereas when targeting IoT, the diversity can be used to ensure robustness and quality of service by increasing the antenna gain of single elements. 
	When employed in the mobile terminal, consequently, the different radiation patterns help to provide an additional link margin. 
	
	IoT applications are expected to be dominated by a large number of cheap devices, requiring low data rates on the one hand.
	On the other hand, IoT applications are expected to require real-time communication and high reliability at the same time.
	As an example, a highly automatized factory environment is discussed in \cite{BLee2018}. 
	
	The requirements of the described factory use case are in contrast to the requirements of ultra-fast Internet and highly demanding data-rate applications. 
	In ultra wideband (UWB) services, a large frequency band is used. LOS communication helps to improve the separation of different data streams, while power consumption is of secondary interest.
	
	Towards the goal of high reliability, a larger link margin by using single-element beamforming is achievable.
	Massive connectivity can be facilitated by boosting the number of data-streams.
	Multi-mode antennas open the flexibility to switch between both options.
	
	As a summary, IoT and UWB applications require different problems to be solved. 
	We propose multi-mode antenna technology to jointly face the requirements. 
	High data rates are offered while reliability and/or connectivity are improved.
	
	Original contributions include: 
	\begin{itemize}
	\item a systematic design procedure for multi-mode antennas using symmetry properties of characteristic modes
	\item a proposal of different multi-mode antenna precoding \mbox{setups} including sophisticated hybrid beamforming structures and single-element beamforming
	\item an evaluation of the achievable gain of single-element beamforming
	\item a sum-rate performance assessment as a function of distance for different multi-mode antenna types
	\end{itemize}
	The remainder is organized as follows. 
	In the next section, the multi-mode antenna design is described. 
	A prototype antenna array providing four ports per element is taken as a reference. 
	New variations are presented, offering up to eight ports on a hexagonally-shaped planar radiator.
	The third section describes the system setup under investigation and introduces different ways of implementation, including aspects on saving hardware costs. 
	Afterwards, the performance of the multi-mode antennas is studied for some example scenarios. 
	Finally, conclusions are drawn.


    \section{Multi-mode Antenna Design}
   
   In order to realize MIMO antennas both at the base station and the mobile terminal, the concept of multi-mode antennas based on the theory of characteristic modes~\cite{Chen2015} is employed. In this concept, different current modes are excited by different antenna ports on a single antenna element. It is thus most suitable for designing MIMO antennas that optimally utilize the given space. Consequently, the concept is purposefully applied to the design of MIMO antennas in spatially restricted environments such as modern mobile devices as well as the design of base station antennas with a compact form factor.
   
   The theory of characteristic modes states that an arbitrary surface current density on a perfectly electrically conducting~(PEC) antenna can be expressed as a weighted sum of characteristic surface current densities. 
   This modal decomposition is described mathematically by a generalized eigenvalue problem, which can be solved numerically by means of the method of moments~(MoM)
   ~\cite{Chen2015}. The solution of the eigenvalue problem yields the characteristic surface current densities and corresponding eigenvalues, which describe how much a characteristic mode may contribute to the total radiation of an antenna.
   
   The characteristic modes possess advantageous orthogonality properties which are leveraged for multi-mode antenna designs. They state that the characteristic far fields are orthogonal to each other, offering pattern and polarization diversity. For this reason, the aim of multi-mode antenna design for MIMO is to respectively excite different sets of characteristic modes with different antenna ports. In this case, the antenna ports are uncorrelated, which is most beneficial for MIMO antennas.

   \subsection{Mobile Terminal Antenna}
   
   \begin{figure}[!t]
   	\centering
   	{\includegraphics[width=3.3in]{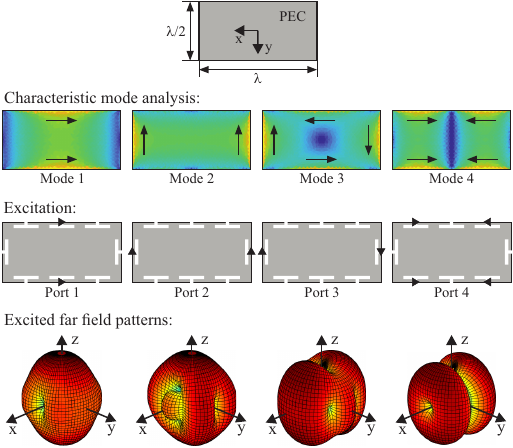}}
   	\caption{Workflow of multi-mode antenna design: A suitable geometry is selected. A characteristic mode analysis is performed yielding significant modes for excitation. Excitations are implemented based on modal currents. $\lambda$: wavelength at design frequency.}
   	\label{fig:CMA}
   \end{figure}
   
   At the mobile terminal, such as a mobile phone or a sensor node, space for housing and placing antennas is usually strictly limited. It is therefore a promising approach to use existing structures like the ground plane of a printed circuit board~(PCB) or the metallic chassis of a device as an antenna. In order to enable MIMO for such antenna structures, the characteristic mode analysis~\cite{Cabedo2007} as described above is a suitable tool. 
   
   Based on the eigenvalues, the characteristic modes which are significant for radiation within the frequency band of interest are identified. For example, the rectangular plate in Fig.~\ref{fig:CMA} with edge lengths of one wavelength times half a wavelength offers four significant characteristic modes, whose surface current densities are depicted. The number of significant modes depends on the electrical size of the antenna.
   
   In order to make use of the diversity potential, excitations have to be defined for the characteristic surface current densities. Typically, two types of excitations are distinguished: Inductive coupling is performed at local maxima of the current density by introducing slots or loops (inductive coupling elements), whereas capacitive coupling is performed at local minima of the current density (local maxima of the electric field) by introducing additional elements to the antenna (capacitive coupling elements)~\cite{Martens2011}. However, it has to be kept in mind that, as the antenna is the chassis itself or an essential part for other functional groups (e.g. a PCB), its shape must not be altered significantly by introducing these coupling elements. Therefore, electrically small coupling elements are typically used. Furthermore, it may be purposeful to drive several such coupling elements simultaneously in order to reproduce a desired current density, depending on the antenna geometry. This concept is also illustrated in Fig.~\ref{fig:CMA}, where four antenna ports are defined by driving different inductive coupling elements simultaneously, inspired by the modal current densities. In such a scenario, a feed network is used which distributes the input signal of an antenna port to the corresponding coupling elements and adjusts the necessary phase and amplitude relations between the coupling elements (see e.g.~\cite{Martens2014}). The effectiveness of a chosen excitation can be analyzed by means of the modal weighting coefficients, which are an additional result of the characteristic mode analysis if excitations are defined. They express to what degree a characteristic mode is excited depending on the similarity of the excitation to the respective characteristic surface current densities and their significance for radiation.
   
   For the design of a MIMO antenna, several antenna ports have to be defined which should be uncorrelated. The port correlation is measured by means of the envelope correlation coefficients. It can be shown that these can also be expressed in terms of the modal weighting coefficients~\cite{Peitzmeier2019}, demonstrating that decorrelation of the antenna ports is achieved if they excite mutually exclusive sets of characteristic modes. This requirement is e.g. fulfilled in Fig.~\ref{fig:CMA}. However, the theory of characteristic modes does not guarantee that the characteristic surface current densities themselves are orthogonal to each other (orthogonality holds for the characteristic far fields, as explained above). On the contrary, the current densities may be correlated, i.e. similar, and thus cannot be excited separately. If different antenna ports excite the same characteristic modes, they will in general be correlated, which should be avoided for MIMO operation. 
   
   In~\cite{Peitzmeier2019}, it is shown that symmetric antennas offer the possibility to realize perfectly uncorrelated antenna ports by utilizing the symmetry properties of characteristic modes (see next subsection). This fact is also made use of in Fig.~\ref{fig:CMA}. However, at the mobile terminal symmetries may not always be exploited since the antenna shape is usually predetermined, as described above. Furthermore, accommodating a feed network for simultaneously driving distributed coupling elements may not be feasible due to spatial restrictions~(cf.~\cite{Martens2014}). 
   
   If a symmetric antenna design is not possible for the above mentioned reasons, an optimization procedure based on the characteristic mode analysis can be employed. It is based on the observation in~\cite{Peitzmeier2019} that certain modal weighting coefficients may be relatively low even for significant modes, which strongly depends on the actual port implementation. Therefore, it is deduced that, if a certain amount of port correlation is allowed, locations for coupling elements can be found that significantly excite only a limited number of characteristic modes. Based on the modal weighting coefficients and their connection to the envelope correlation coefficients, the antenna ports are optimized in such a way that the significantly excited characteristic modes per port are distinct and only the weakly excited modes overlap. This way, weakly correlated antenna ports can be realized iteratively in an automated manner. As the proposed procedure is based on modal parameters alone, only one full simulation run of the given antenna geometry is needed for getting the modal results. Simulations of the complete system yield that the maximum allowed correlation should not exceed~\SI{-9}{\decibel} (see the section on system evaluation).

   \subsection{Base Station Antenna}
   
   \begin{figure}[!t]
   	\centering
   	{\includegraphics[width=3.3in]{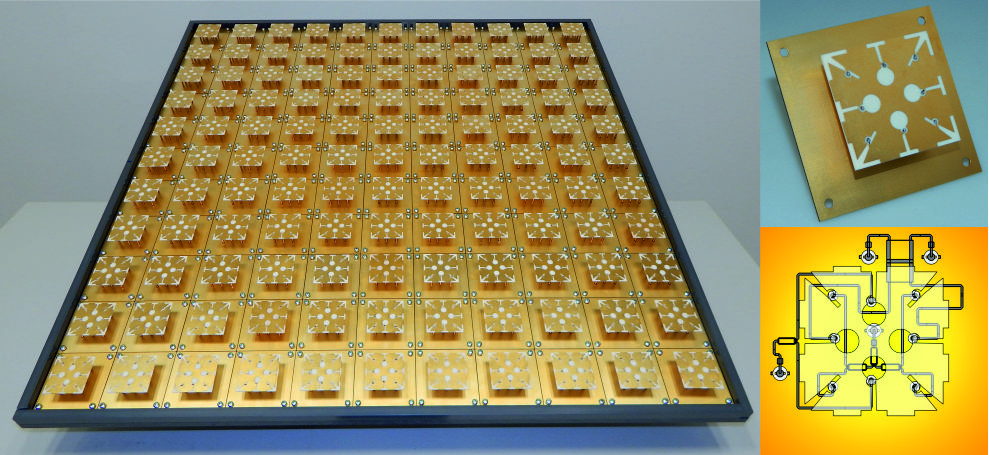}}
   	\caption{Multi-mode massive MIMO antenna array as designed in~\cite{Manteuffel2016} consisting of~$11\times 11$ four-port antenna elements (see insets) based on a square geometry with feed networks using power dividers and hybrid couplers realized in stripline technology.}
   	\label{fig:M4_antenna}
   \end{figure}
   
   At the base station, a massive MIMO antenna array is employed. In order to utilize the given space more efficiently than conventional arrays, multi-mode antennas are used as array elements. These elements can be arranged in the same way as in conventional arrays, e.g. as a two-dimensional linear array, yielding a multi-mode massive MIMO antenna array~\cite{Manteuffel2016}.
   
   In contrast to the mobile terminal, the elements of the base station antenna array can be designed freely. The goal is to design compact antenna elements with as many uncorrelated ports as possible. Recent results show that symmetric antenna elements are particularly suitable to achieve this goal~\cite{Peitzmeier2019}: The characteristic surface current densities of a symmetric antenna can be sorted into a limited number of sets which are mutually orthogonal, depending on their symmetry properties. Choosing symmetrically placed excitations, these sets of characteristic modes can be excited separately, yielding perfectly uncorrelated antenna ports. However, as the number of symmetry operations of an antenna is limited, the number of mutually orthogonal sets of characteristic surface current densities is limited, i.e. there is an upper bound for realizing uncorrelated antenna ports which depends on the symmetry order of the antenna geometry.
   
   For example, a square plate offers six sets of mutually orthogonal characteristic surface current densities and thus six uncorrelated antenna ports, as demonstrated in~\cite{Peitzmeier2019}. Furthermore, a square plate is well suited for arrangement in an array due to its geometry. It should be noted, however, that fulfilling the symmetry requirements for realizing uncorrelated antenna ports may result in a comparatively complex feed network (cf.~\cite{Peitzmeier2019}). If the port potential is not fully exploited, the feed network becomes less complex and the antenna size can be further reduced, as shown in~\cite{Manteuffel2016}, where a complete prototype of a multi-mode massive MIMO array consisting of four-port antenna elements based on a square plate is presented~(Fig.~\ref{fig:M4_antenna}).
   
   The prototype in~\cite{Manteuffel2016} also illustrates how the excitations for such an antenna element can be realized. As the antenna shape can be designed freely, the exciters do not have to be electrically small, in contrast to the mobile terminal, allowing inherent impedance matching by optimizing the geometry of the exciters. This way, input reflection coefficients of less than~\SI{-10}{\decibel} are achieved from~\SI{6}{\giga\hertz} to~\SI{8.5}{\giga\hertz}. It is important, however, that the overall symmetry of the antenna must not be altered in order to ensure uncorrelated antenna ports~\cite{Peitzmeier2019}. Therefore, the exciters have to be placed symmetrically, as visible in Fig.~\ref{fig:M4_antenna}. 
   
   Furthermore, a feed network is needed in order to distribute the input signals at the antenna ports to the corresponding exciters on the antenna element with the correct amplitude and phase relations. In this case, a feed network consisting of Wilkinson power dividers and a \SI{90}{\degree}-hybrid~coupler is realized in stripline technology~(Fig.~\ref{fig:M4_antenna}). The mutual coupling between the ports of a single antenna element is very low due to its symmetric design~\cite{Peitzmeier2019}. Thus, the only source for coupling is the feed network, which should therefore be designed for high isolation. The prototype in~\cite{Manteuffel2016} achieves coupling coefficients of less than~\SI{-20}{\decibel} per antenna element.
   
   In order to realize even more antenna ports per array element, geometries with a higher symmetry order should be employed. An interesting geometry is the hexagonal plate, which provides up to eight uncorrelated antenna ports~\cite{Peitzmeier2019_EuCAP}. Moreover, it offers the possibility for a new arrangement of the antenna elements within the array, i.e. a hexagonal tiling. It should be noted that, as the number of antenna ports grows, the complexity of the feed network increases~(cf.~\cite{Martens2014, Peitzmeier2019, Manteuffel2016}), potentially becoming a limiting factor for practical designs.
   
   For all the aforementioned geometries, the antenna size can be minimized in such a way that only the necessary modes for achieving the maximum number of orthogonal ports are significant. Based on the antenna symmetry, the necessary modes can be determined and the antenna size can be set such that only these modes are significant. For the above-mentioned square plate with four ports, square plate with six ports, and the hexagonal plate with eight ports, it is found experimentally that the minimum circumference radius is approximately 0.6~wavelengths, 0.65~wavelengths, and 0.7~wavelengths, respectively, at the design frequency.
   
   \begin{table}[!t]
   	{\color{black}
   		\renewcommand{\arraystretch}{1.3}
   		\caption{Parameters of Multi-Mode Massive MIMO Arrays ($\lambda$:~Wavelength at Design Frequency)}
   		\label{tab:Array}
   		\centering
   		\begin{tabular}{|p{1.1in}|p{0.85in}|p{0.8in}|}
   			\hline
   			Array/Antenna type & Four-port multi-mode antennas & Crossed dipoles \\
   			\hline
   			Inter-element coupl. & -25~DB & -25~DB  \\
   			\hline
   			Number of ports & 484 & 484 \\
   			\hline
   			Ports per element & 4 & 2 \\
   			\hline
   			Element size & 0.72$\lambda^2$ & 0.25$\lambda^2$ \\
   			\hline
   			Inter-element spacing & 0.58$\lambda$ & $\lambda$ \\
   			\hline
   			Total array size & 247$\lambda^2$ & 544$\lambda^2$ \\
   			\hline
   	\end{tabular}}
   \end{table}
   As the multi-mode antenna elements are electrically large, they should be placed as close as possible in order to realize a compact array. Experimental results in~\cite{Manteuffel2016} show that coupling coefficients between elements of less than~\SI{-25}{\decibel} can be achieved with inter-element spacings of approximately 0.6~wavelengths and the employment of choke walls. Compared to a generic array of crossed dipoles achieving the same decoupling without further measures, this results in an overall size reduction of 54~percent. Table~\ref{tab:Array} summarizes the important array parameters.


	\section{Baseband System Architecture}
	
	Based on the antenna characteristics under investigation, in this section the baseband system architecture and possible adaptations thereof are described.
	In Fig. \ref{fig: BSB_analog_pattern}, a block diagram of the transmitter is depicted.
	A multicarrier (OFDM-like) system is assumed.
	The input data vectors are processed by digital precoding matrices.
	The digital precoding matrices are chosen individually for every subcarrier.
	The outputs of the digital precoding are connected by means of radio frequency (RF) chains to an analog precoder.
	An RF chain includes the digital to analog conversion, filtering and mixing of the signals to the desired frequency band.
	It is assumed that the hardware used in the analog precoder is wideband.
	
	\begin{figure}[h]
		\centering
		\def\svgwidth{0.9\columnwidth}
		\input{./pic/BSB_analog_pattern_New_Names.pdf_tex}
		\caption{Block diagram of the transmitter. I) denotes the number of available RF chains, II) the number of antenna elements, and III) the number of ports at the multi-mode antenna element, respectively.}
		\label{fig: BSB_analog_pattern}
	\end{figure}
	\subsection{\textcolor{black}{Digital Beamforming}}
	In the special case of digital beamforming, the analog precoder provides a direct connection from a single antenna port to a single RF chain. Correspondingly, the number of RF chains is identical with the number of transmit antenna ports, which is defined by the number of antennas and the number of ports per antenna.
	In conjunction with massive MIMO systems, both hardware cost and power consumption are excessive.
	Digital beamforming offers the largest freedom and hence provides best system performance.
	The performance of digital beamforming provides a reference.
	When the maximum tolerable correlation between two antenna ports shall be evaluated, a Kronecker model can be used.
	The loss of achievable sum-rate serves as a measure of the amount of tolerable correlation.
	
	\subsection{Approaches to Hybrid Beamforming}

	Besides digital beamforming and a fully connected approach, where each antenna port is connected to each RF chain, sub-connected architectures are feasible. 
	
	To reduce the hardware complexity, user-scheduling algorithms can be used.
	Numerous user and stream allocation techniques both for digital beamforming as well as hybrid beamforming have been published.
	However, most hybrid user scheduling algorithms are determined and modified from counterparts used for digital beamforming.
	Commonly, in hybrid beamforming schemes, the degrees of freedom are less than in digital beamforming.
	This limitation is due to a limited number of RF chains.
	An additional limitation is caused by the analog beamforming, since hardware components are not as flexible as software is. 
	Furthermore, the characteristics of the employed hardware need to be known for optimization purposes.
	This requires some additional channel estimation and measurement effort.
	Further details on different beamforming techniques can be found in \cite{Ahmed2018}.
	
	For reasons of complexity, subsequently focus is on hybrid beamforming. 
	Specifically, the following methods are studied:
	\begin{itemize}
		\item fully-connected approach 
		\item spatial-filtering approach
		\item single-element beamforming approach
	\end{itemize}
	\subsubsection{Fully-Connected Approach}
	In the fully-connected approach, each RF chain is connected to each antenna port by means of steerable phase shifters and matching networks. Hence, the most flexible form of hybrid beamforming is a fully connected architecture.
	As this requires additional hardware devices and therefore increases costs while reducing power efficiency, sub-array structures are of interest.
	When thinking about multi-mode antennas, two setups based on sub-array precoding are intuitive.
	\subsubsection{Spatial Filtering Approach}
	Given an array based on identical multi-mode antenna elements, in the first setup all ports employing the same set of modes are connected to form a sub-array. Recall that the RF chains are only connected to a part of the available ports.
	The spatial characteristics of the structure are used to increase systems performance. 
	This allows simple calculation of the array factor, since simple uniform linear/planar array (ULA/UPA) processing can be used if the multi-mode antenna elements are uniformly spaced. 
	Ideally, the array characteristics are employed in order to suppress undesired sidelobes of the chosen group of modes.
	Therefore, this technique is referred to as spatial filtering.
	
	\subsubsection{Single-Element Beamforming Approach}
	The second possibility is to switch between all ports of a single radiation element, subsequently using different characteristics at different positions in the overall sub-array. 
	Thus, mode and space in combination can be used in order to optimize reception and radiation.
	The single-element beamforming approach requires more sophisticated calculations \cite{Doose2018}, since the different radiation patterns need to be included.
	The pattern corresponding to one group of modes of the \mbox{(sub-)} array is only achieved, if only one port of the antenna element is used.
	The radiation pattern of the port fitting best to the desired direction needs to be chosen. This is referred to as mode-selection.
	
	A more sophisticated method is to combine the individual groups of modes belonging to the ports of a single element.
	Here, the sub-arrays interfere not only in terms of spatially founded interference, but additionally in terms of mode diversity.
	The before mentioned sub-arrays holding the same radiation patterns still are employed.
	The different modes can be combined such that the radiation given a desired direction is optimized.
	If only one element is used for this kind of beamforming, this is dubbed single-element beamforming subsequently.
	Each radiating element of a multi-mode antenna array is treated as a sub-array here. 
	By using a priori knowledge of the individual modes, the modes can be combined in a way such that the radiation of a single antenna element is steered towards a single angle.
	This enables ULA/UPA processing while using the individual antenna elements as a smart antenna.
	One drawback is that additional channel information is required.


	\section{System Evaluation}
	
	In this section different results on the requirements of IoT and UWB Internet access are provided and their application is explained.
	In UWB systems, high sum-rates are targeted. 
	Contrary, in IoT applications, high reliability and/or massive connectivity are desirable.
	Multi-mode antennas open the flexibility to provide a trade-off.
	As shown before, complex but symmetric shapes can be used in order to implement multiple ports on a given surface. 
	Using the radiation characteristics of the prototype antenna in Fig. \ref{fig:M4_antenna} as an example, a single element beamforming approach is presented.
	In the block diagram presented in Fig.~\ref{fig: BSB_analog_pattern}, on the right-hand side a cut of the radiation pattern in the y,z-plane is depicted.
	Cuts of the patterns are redrawn in Fig.~\ref{fig: Single_Element_Beamforming}, curves (1)-(4).
	Since the ports corresponding to the electric fields are orthogonal among each other, single-element beamforming can be used to increase the overall antenna gain given a certain direction.
	To achieve a fair comparison, the input power is kept constant.
	At the mobile terminal this is assumed as an alternative to simplify the problem structure and provide a minimum channel performance at the cost of parallel data streams.
	The achievable gain of mode combination is represented by mode-combination in Fig. \ref{fig: Single_Element_Beamforming}, curve (6).
	It is shown that by combining the ports of the antenna and using constructive interference, an additional gain of up to two decibel can be achieved.
	Note that digital beamforming at the antenna terminal is required to achieve the full performance.
	A codebook is determined, storing the optimal amplitudes and phases.
	A still attractive performance because of reduced complexity is switching to the mode providing the largest gain in the desired direction (mode-selection, Fig. \ref{fig: Single_Element_Beamforming}, curve (5)).
	Here, a gain of about 6~DBi can be achieved for nearly the full angular range.
	The corresponding codebook only stores the selected mode, which can be set by using switches, as an example.
	
	\input{./pic/DrawSingleElementBeamforming.tex}
	
	In \cite{Hoeher2017}, the sum-rate is plotted as a function of distance between the base station array presented in \cite{Manteuffel2016} and a mobile station.
	The work is extended here by using the results of the analysis presented above.
	One user employing the same type of antenna as used in the BS is assumed.
	The BS is assumed to provide a fixed number of 121 radiators.
	To allow zero-forcing (ZF) precoding, digital beamforming is employed and the range estimation is conducted based on the WINNER II channel model, scenario A1, NLOS \cite{WINNER2008}.
	A fair comparison is achieved by using the same signal-to-noise ratio settings, corresponding to the total transmit power.
	As can be seen, both data rate and achievable range can be increased. 
	Note that for simple receiver design ZF precoding is applied.
	For better comparison of the antennas, at the MS the same antenna type as in the BS is assumed.
	
	\input{./pic/DrawRangeEstimation.tex}
	
	Although the realization of highly uncorrelated ports is theoretically possible, this requires an improved hardware effort, as discussed above and in \cite{Peitzmeier2019,Manteuffel2016}. The described Kronecker channel model is employed to evaluate the impact of sub-optimal orthogonality. A correlation power of roughly -12~DB is found to be tolerable. Since the correlation occurs twice, at receive and transmit side, 3~DB can be added, which results in a correlation of -9~DB.
	

	\section{Conclusions}
	
	In this contribution, multi-mode antennas are studied to enable ultra-high data rates of 100~Gb/s and beyond in the frequency region below 10~GHz, while increasing the reliability of service.
	The workflow based on symmetry considerations is proposed as a suitable tool for designing multi-mode antennas with a high number of orthogonal ports.
	Different beamforming system designs matched to multi-mode antennas are suggested and their differences are determined.
	The sum-rate performance is evaluated as a function of the distance between a multi-mode BS antenna array and a mobile station equipped with a single multi-mode antenna.
	The influence of the number of available ports is studied with respect to data rate and distance between BS and mobile terminals. 
	Single-element beamforming is shown to be particularly useful in IoT applications, whereas the multi-mode antenna array architecture at the BS is a key enabling technique for ultra-wideband Internet access and related applications.

	\section*{Acknowledgment}
	
	This work has been funded within the priority program DFG SPP 1655 conducted by the German Research Foundation.
	


\section*{Biographies}
	
	\textsc{Nils Lennart Johannsen} (nj@tf.uni-kiel.de) received his B.Eng. from Hamburg University of Applied Sciences, Hamburg, Germany, in 2016, and an M.Sc. degree in electrical engineering and information technology from Kiel University, Kiel, Germany, in 2018. He is currently pursuing the Ph.D. degree with the Chair of Information and Coding Theory, Kiel University, as a Research and Teaching Assistant. His research interests include massive MIMO systems and baseband signal processing for multi-mode antennas.\\
	
	\textsc{Nikolai Peitzmeier} (peitzmeier@hft.uni-hannover.de) received his B.Sc. and M.Sc. degrees in electrical engineering and information technology from the Leibniz University Hannover, Hannover, Germany, in 2013 and 2014, respectively. Since then, he has been a Research Assistant with the Institute of Microwave and Wireless Systems, Leibniz University Hannover. His current research is focused on multiple-input multiple-output (MIMO) antenna systems based on the theory of characteristic modes, including beamforming and massive MIMO approaches.\\
	
	\textsc{Peter Adam Hoeher} [F '14] (ph@tf.uni-kiel.de) received the \mbox{Dipl.Ing.} degree from RWTH Aachen University, Germany, in 1986 and the \mbox{Dr.Ing.} degree from the University of Kaiserslautern in 1990, both in electrical engineering. From 1986 to 1998, he was with the German Aerospace Center, Oberpfaffenhofen. From 1991 to 1992, he was on leave at AT\&T Bell Laboratories, Murray Hill, NJ. Since 1998 he is a Full Professor of electrical and information engineering at Kiel University, Germany. \\
	
	\textsc{Dirk Manteuffel} (manteuffel@hft.uni-hannover.de) received the \mbox{Dipl.Ing.} and \mbox{Dr.Ing.} degrees in electrical engineering from the University of Duisburg-Essen, Essen, Germany, in 1998 and 2002, respectively. From 1998 to 2009, he was with IMST, Kamp-Lintfort, Germany. From 2009 to 2016, he was a Full Professor of wireless communications at Christian-Albrechts-University, Kiel, Germany. Since June 2016, he is a Full Professor and the Director of the Institute of Microwave and Wireless Systems, Leibniz University Hannover, Hannover, Germany.

\end{document}

%% file: pic/BSB_analog_pattern_New_Names.pdf_tex
\begingroup%
  \makeatletter%
  \providecommand\color[2][]{%
    \errmessage{(Inkscape) Color is used for the text in Inkscape, but the package 'color.sty' is not loaded}%
    \renewcommand\color[2][]{}%
  }%
  \providecommand\transparent[1]{%
    \errmessage{(Inkscape) Transparency is used (non-zero) for the text in Inkscape, but the package 'transparent.sty' is not loaded}%
    \renewcommand\transparent[1]{}%
  }%
  \providecommand\rotatebox[2]{#2}%
  \ifx\svgwidth\undefined%
    \setlength{\unitlength}{871.52810885bp}%
    \ifx\svgscale\undefined%
      \relax%
    \else%
      \setlength{\unitlength}{\unitlength * \real{\svgscale}}%
    \fi%
  \else%
    \setlength{\unitlength}{\svgwidth}%
  \fi%
  \global\let\svgwidth\undefined%
  \global\let\svgscale\undefined%
  \makeatother%
  \begin{picture}(1,0.60805964)%
    \put(0,0){\includegraphics[width=\unitlength,page=1]{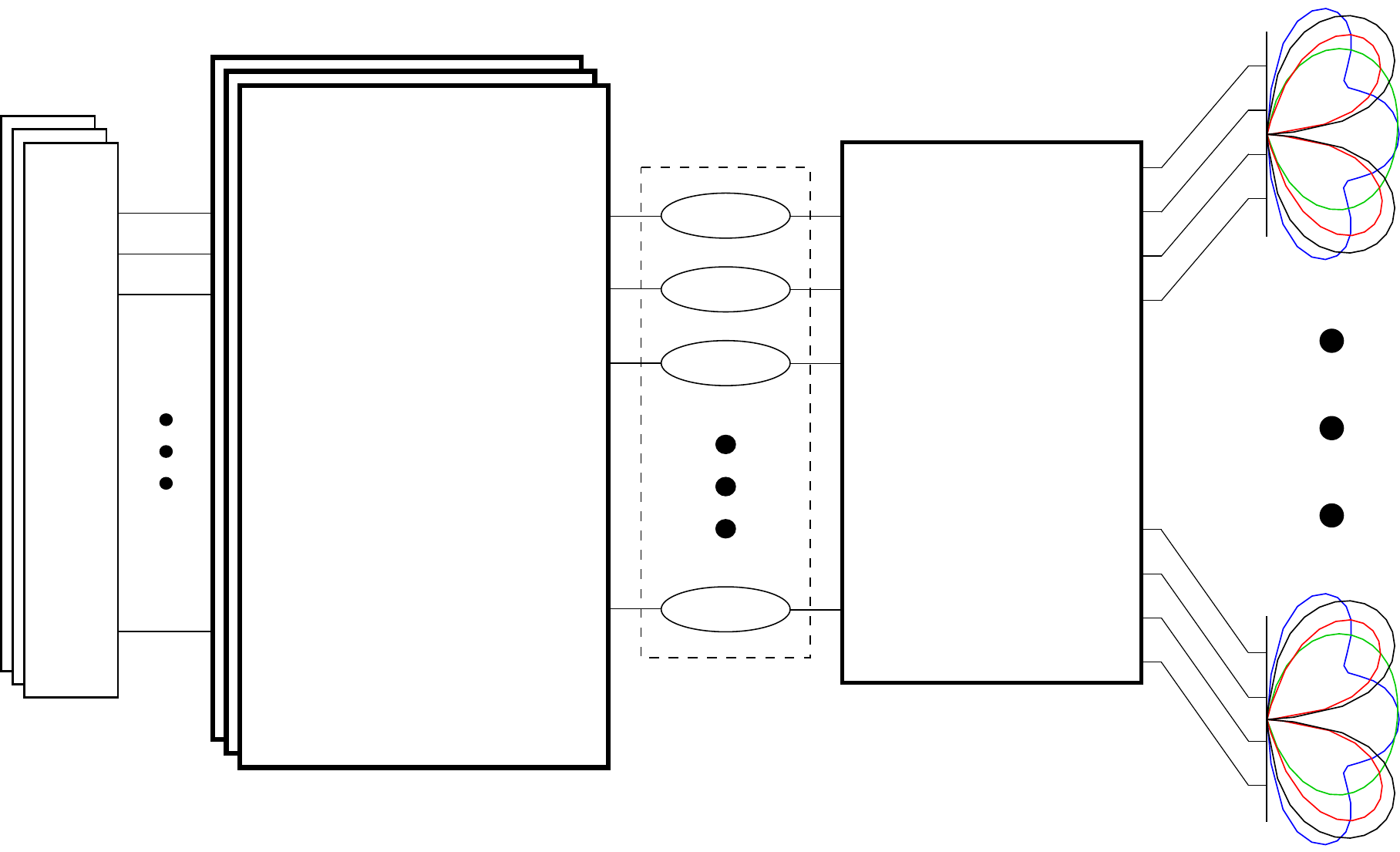}}%
    \put(0.17871906,0.36202689){\color[rgb]{0,0,0}\makebox(0,0)[lt]{\begin{minipage}{0.34008745\unitlength}\raggedright Digital\\ Precoders\end{minipage}}}%
    \put(0.60898917,0.36077472){\color[rgb]{0,0,0}\makebox(0,0)[lt]{\begin{minipage}{0.38365017\unitlength}\raggedright Analog\\ Precoder\end{minipage}}}%
    \put(0.46593042,0.50214257){\color[rgb]{0,0,0}\makebox(0,0)[lb]{\smash{I)}}}%
    \put(0,0){\includegraphics[width=\unitlength,page=2]{./pic/BSB_analog_pattern_New_Names_.pdf}}%
    \put(0.87597056,0.39064003){\color[rgb]{0,0,0}\makebox(0,0)[lb]{\smash{III)}}}%
    \put(0.06184514,0.17147015){\color[rgb]{0,0,0}\rotatebox{90}{\makebox(0,0)[lb]{\smash{Data Vectors}}}}%
    \put(0,0){\includegraphics[width=\unitlength,page=3]{./pic/BSB_analog_pattern_New_Names_.pdf}}%
    \put(0.87597056,-0.02792229){\color[rgb]{0,0,0}\makebox(0,0)[lb]{\smash{III)}}}%
    \put(0,0){\includegraphics[width=\unitlength,page=4]{./pic/BSB_analog_pattern_New_Names_.pdf}}%
    \put(0.87597056,0.61948242){\color[rgb]{0,0,0}\makebox(0,0)[lb]{\smash{II)}}}%
  \end{picture}%
\endgroup%

%% file: pic/DrawSingleElementBeamforming.tex
\begin{figure}[h]
	\centering
	\begin{tikzpicture}
		\begin{axis}[
		legend columns = 6,
		legend style={at={(0.4,-0.23)},anchor=north},
		legend style = {draw,fill=white},
		xtick = {-90,-45,...,90},
		legend cell align = left,
		width = 0.95\columnwidth,grid = both,
		xmin = -90, xmax = 90, ymin = 0, ymax = 11,
		y tick label style={/pgf/number format/1000 sep=.},
		ylabel = Achieved {gain} in ${{\textrm{DBi}}}$,
		xlabel = Angle in degree,
		table/x index={0},
		x filter/.code=\pgfmathparse{#1-180}
		]

		\addplot[blue, mark = x] table[y index = {1}, col sep = comma]{./pic/E_CWC_NM4_Rearranged_for_MaxEIRP.csv};
		\addplot[green!80!black, mark = diamond] table[y index = {2}, col sep = comma]{./pic/E_CWC_NM4_Rearranged_for_MaxEIRP.csv};
		\addplot[red, mark = x] table[y index = {3}, col sep = comma]{./pic/E_CWC_NM4_Rearranged_for_MaxEIRP.csv};
		\addplot[black, mark = diamond] table[y index = {4}, col sep = comma]{./pic/E_CWC_NM4_Rearranged_for_MaxEIRP.csv};
		\addplot[orange!80!black, mark = triangle,thick] table[y index = {1}, col sep = comma]{./pic/MaxEIRP4ModesdB.csv};
		\addplot[black, mark = o] table[y index = {1}, col sep = comma]{./pic/20190127_EIRP_4modes_t_dB.csv};
		
		\legend{\small{(1)},\small{(2)},\small{(3)},\small{(4)},\small{(5)},\small{(6)}}
		\end{axis}
	\end{tikzpicture}
	\caption{
		Maximum achievable gain of a single element per angle. Available modes are plotted for reference (1)-(4). By choosing the best mode at a certain angle (5) is achieved. The performance can be improved significantly when certain modes are combined (6).}
	\label{fig: Single_Element_Beamforming}
	
\end{figure}

%% file: pic/DrawRangeEstimation.tex
\begin{figure}[h]
	\centering
	\begin{tikzpicture}
		\begin{axis}[
		legend pos = north east,
		legend columns = 3,
		legend style = {draw,fill=white},
		legend cell align = left,
		width = 0.95\columnwidth,grid = both,
		xmin = 1, xmax = 10, ymin = 0, ymax = 200,
		y tick label style={/pgf/number format/1000 sep=.},
		ylabel = Sum-rate in ${{\textrm{Gb/s}}}$,
		xlabel = Distance between BS and user in m]
		
		\addplot[black, mark = diamond] table[x expr= \thisrow{D},y expr= \thisrow{R}/1e9, col sep = tab]{./pic/ict-sim-m4-link-budget-rates-vs-distance_m8.dat};
		\addplot[black, dashed, every mark/.append style={solid}, mark = o] table[x expr= \thisrow{D},y expr= \thisrow{R}/1e9, col sep = tab]{./pic/ict-sim-m4-link-budget-rates-vs-distance_m6.dat};
		\addplot[black,dotted, every mark/.append style={solid}, mark = x] table[x expr= (\thisrow{D}),y expr = \thisrow{R}/1e9, col sep = tab]{./pic/ict-sim-m4-link-budget-rates-vs-distance_m4.dat};
	
		\addlegendentry{\small{8 ports}}
		\addlegendentry{\small{6 ports}}
		\addlegendentry{\small{4 ports}}
		
		\end{axis}
	\end{tikzpicture}
	\caption{Expected overall info bit rate as a function of distance extending the work in \cite{Hoeher2017}. The assumptions are as follows: The BS has 121 elements, at the receiver one element is employed. A 968x8 (8 ports per element), 726x6 (6 ports per element) and 484x4 (4 ports per element) massive MIMO system in downlink scenario is specified. The channel is generated according to the WINNER II A1 NLOS scenario. The coding scheme according to the DVB-S2 standard with bit rates of \added{ $1/4$ to $9/10$} and QAM modulation orders of \added{$4$ to $16384$} are used. The target bit error rate is assumed to be $10^{-6}$.}
	\label{fig: DrawRangeEstimation}
	
\end{figure}

%% file: ICT_HFT_M4_CommunicationsMagazine_Final_OneFile.bbl
\begin{thebibliography}{99}
		
		\bibitem{Militano2015}
		L.~Militano {\em et al.}, 
		``Device-to-device communications for {5G} internet-of-things,'' {\em EAI Endorsed Trans. on Internet Things}, vol. 15, Oct. 2015.
		
		\bibitem{Ejaz2016}
		W.~{Ejaz} {\em et al.}, 
		``Internet of things ({IoT}) in {5G} wireless communications,'' {\em {IEEE Access}}, vol.~4, pp.~10310--10314, 2016.
		
		\bibitem{BLee2018}
		B.M.~Lee,
		``Improved energy efficiency of massive MIMO-OFDM in battery-limited {IoT} networks'', {\em IEEE Access}, vol. 6, pp. 38147-38160, June 2018.
		
		\bibitem{Chen2015}
		Yikai~Chen and Chao-Fu~Wang,
		{\em Characteristic Modes: Theory and Applications in Antenna Engineering}, Hoboken, New Jersey,
		Wiley, 2015.
		
		\bibitem{Cabedo2007}
		M.~Cabedo-Fabres {\em et al.}, 
		 ``The theory of characteristic modes revisited: a contribution to the design of antennas for modern applications,'' in {\em IEEE Antennas Propag. Mag.}, vol.~49, no.~5, pp.~52-68, Oct.~2007.
		
		\bibitem{Martens2011}
		R.~Martens, E.~Safin and D.~Manteuffel,
		``Inductive and capacitive excitation of the characteristic modes of small terminals,''
		in {\em 2011 Loughborough Antennas \& Propag. Conf.}, Loughborough, UK, Nov.~2011.
		
		\bibitem{Martens2014}
		R.~Martens and D.~Manteuffel,
		``Systematic design method of a mobile multiple antenna system using the theory of characteristic modes,''
		{\em IET Microwaves, Antennas \& Propag.}, vol.~8, no.~12, pp.~887--893, Sept.~2014.
		
		\bibitem{Peitzmeier2019}
		N.~Peitzmeier and D.~Manteuffel,
		``Upper bounds and design guidelines for realizing uncorrelated ports on multi-mode antennas based on symmetry analysis of characteristic modes,''
		{\em IEEE Trans. Antennas Propag.}, vol.~67, no.~6, pp.~3902-3914, June~2019.
		
		\bibitem{Manteuffel2016}
		D.~Manteuffel and R.~Martens,
		``Compact multimode multielement antenna for indoor UWB massive MIMO,''
		{\em IEEE Trans. Antennas Propag.}, vol.~64, no.~7, pp.~2689-2697, July~2016.
		
		\bibitem{Peitzmeier2019_EuCAP}
		N.~Peitzmeier and D.~Manteuffel,
		``Multi-mode antenna concept based on symmetry analysis of characteristic modes,''
		in {\em 13th Europ. Conf. Antennas Propag. (EuCAP 2019)}, Krakow, Poland, Apr.~2019.
		
		\bibitem{Ahmed2018}
		I.~Ahmed {\em et al.}, ``A survey on hybrid beamforming techniques in 5G{:} architecture and system model perspectives,'' {\em IEEE Commun. Surveys Tuts.}, vol. 20, no. 4, pp. 3060–3097, 2018.
		
		\bibitem{Doose2018}
		N.~Doose and P.A.~Hoeher,
		``Joint precoding and power control for {EIRP}-limited {MIMO} systems,''
		{\em IEEE Trans. Wireless Commun.}, vol. 17, no. 3, pp. 1727-1737, Mar. 2018.
		
		\bibitem{WINNER2008}
		P. Kyösti {\em et al.}, 
		``WINNER II channel models,'' 02 2008, IST-4-027756 WINNER II D1.1.2 V1.2.
		
		\bibitem{Hoeher2017}
		P.A. Hoeher {\em et al.}, 
		 ``Ultra-wideband massive MIMO communications using multi-mode antennas,''
		{\em Frequenz}, vol. 71, no. 9-10, pp. 429–448, Sept. 2017.
		
		
		
		
	\end{thebibliography}
